\documentclass[a4paper,12pt]{article}

\usepackage{graphicx}
\usepackage{dcolumn}
\usepackage{bm}
\usepackage{verbatim}
\usepackage{amsmath,amssymb}
\usepackage{relsize,exscale}
\usepackage{color}
\usepackage{hyperref}
\begin{document}

\title{ Relations between different notions of degrees of freedom of a quantum system and its classical model}
 \author{
 N. Buri\' c\thanks{e-mail: buric@ipb.ac.rs}
 \\  Institute of Physics, University of Belgrade,\\ PO BOX 68, 11000 Belgrade,
 Serbia.}

 \maketitle

\begin{abstract}

There are at least three different notions of degrees of freedom (DF) that are important in comparison of quantum and classical dynamical systems.
One is related to the type of dynamical equations and inequivalent initial conditions, the other to the structure of the system and the third
 to the properties of dynamical orbits. In this paper, definitions and comparison in classical and quantum systems of the tree types of DF are
  formulated and discussed.
In particular, we concentrate on comparison of the number of the so called dynamical DF in a quantum system and its classical model. The comparison involves analyzes of relations between integrability of the classical model, dynamical symmetry and separability of the quantum and the
 corresponding classical systems and dynamical generation of appropriately defined quantumness.
   The analyzes is conducted using illustrative typical systems.
  A conjecture summarizing the observed relation between
    generation of quantumness by the quantum dynamics and dynamical properties of the classical model is formulated.

\end{abstract}


\newpage

\section{Introduction}

Relation between quantum  and the corresponding classical systems is indeed a complex one \cite{Landsman}.
The differences between the two theories have induced major changes in our understanding of Reality \cite{dEsp} and have been used in major technological developments \cite{tech}.
 Perhaps equally interesting  are various similarities or analogies between quantum and classical theories \cite{corr}. Such notions, with the
  apparently similar meaning and role in quantum and in classical mechanics, are notions of number and type of degrees of freedom that are
   relevant for the description of a system's structure.  However, there are at least three different concepts that can be justifiably called degrees of freedom. One of these is rather formal and mathematical since it is determined by the mathematical nature of the relevant evolution equation.
 Since the type of evolution equations of quantum mechanical systems,  i.e. partial differential equation,  is radically different from that of the corresponding classical mechanical systems, i.e. ordinary differential equations, the number of the degrees of freedom in this sense is also different. The second definition concentrates on an abstract definition of the
  system's structure as formalized by the Lie algebra of distinguished dynamical variables. The structure of a quantum system and the corresponding classical system are characterized by the same Lie algebra, and the two system have the same number and type of the correspondingly defined degrees
   of freedom. The third notion of the number of degrees of freedom that we shall introduced and call the dynamical degrees of freedom, is fundamentally related to the dynamics displayed
    by a particular system.
   We shall see that the relation between the number of degrees of freedom in this third sense for a quantum and the corresponding classical system
    is rather nontrivial. In order to study this relation we shall need to use an appropriate notion of quantumness of a quantum state, and to compare dynamical changes of the quantumness with the properties of dynamics of the corresponding classical model. To this end, we shall numerically
     analyze quantum and classical dynamics for the examples  of two typical systems. This analyzes will lead us to formulate our major conjecture concerning the relation between quantumness generation, dynamical properties  and the numbers of dynamical degrees of freedom of a quantum system and its classical model.

In the next section we shall discuss the three notions of degrees of freedom.
Algebraic definition of the structural degrees of freedom  was
introduced for quantum systems in references \cite{Cheng1,Cheng2,Cheng3}. We have slightly generalized the presentation so that it automatically applies on classical Hamiltonian systems as well as on the quantum systems. In this section we introduce and stress the importance of the third type, so called the dynamical degrees of freedom.
The quantumness of states with respect to the selected set of the basic variables is introduced in section 3, and is similar in spirit with the notion of generalized entanglement presented
in \cite{Viola1,Viola2,Viola3},  and similarly
 in \cite{Klyasko} and \cite{Zanardi}.
In section 4. we present the notion of a classical model as a constrained
 quantum system such that the quantumness is preserved minimal during the evolution. This way of defining the classical models was introduced in \cite{usPRA1} for a system of oscillators
  and in \cite{usPRA2} for spin systems. Section 5 treats examples in order to infer possible relations between the numbers of dynamical degrees of
   freedom for a quantum system and its classical model. The results suggest our major conjecture formulated in section 5.2.
   Here presented view of
the general relation between the number of dynamical DF for quantum and the corresponding classical system and the  dynamics of
classical models of quantum systems and that of the
generalized quantumness has not been discussed before. Summary is given in section 6.

\section{Three types of degrees of freedom}

We shall argue that it makes sense to introduce at least three different types of degrees of freedom corresponding to different intuitions about what
a notion of degrees of freedom should mean. The three types will be called: a) formal DF (FDF); b) structural or algebraic DF (SDF) and c) dynamical DF (DDF). The first corresponds to the dimensionality of the manifold of inequivalent initial conditions and is thus related to the type of dynamical equations.
 The second is related to the structural properties of the system and is formulated using the {Lie} algebraic relations between the basic physical
  variables of the system. In fact, SDF become specified as a crucial part of the definition of the system under consideration. The Hamiltonian is also involved in this considerations, but without any reference to the system's dynamics. In fact it is only the kinematical part of the Hamiltonian
  that is relevant for SDF.
   The third notion, which we call dynamical DF (DDF), is related to the dimension of typical invariant manifold of the dynamics. The most commonly
    used notion of degrees of freedom is perhaps closest to the notion of SDF. Together with FDF, SDF and DDF we shall use the notation
    NFDF, NSDF and NDDF to denote the numbers of the corresponding DF.

 \subsection{ Formal degrees of freedom}

 Let us first define the notion of FDF. The  {\it type} of a dynamical system under consideration is defined by the set of evolution or dynamical equations of first order in the evolution parameter. The Cauchy problem for different types of  dynamical equations requires different phase spaces or manifolds of initial conditions. Hamiltonian mechanical system is formulated on
  a 2N-dimensional real manifold, evolution equation of a field $\psi(x), x\in {\mathbf R}$, like for example the Schr\" odinger equation, is formulated on an infinite space of functions, and the evolution of an $N$ level quantum systems is formulated on an $N$-dimensional complex space.

 {\it Definition of FDF: The number of FDF of a dynamical system is defined as half the number of the real dimensions of the manifold of physically inequivalent initial conditions for the relevant evolution equations.}

  For a Hamiltonian system with a $2N$-dimensional real phase space ($N$ might be $\infty$ as for fields), and this also covers the quantum unitary evolution (please see the {\bf Remark 1}) the number of FDF is $N$.

  The specification of "physically inequivalent initial conditions"  refers to the general theory and depends on the type of dynamical equations or in other words on the constraints imposed by the type of
   dynamics and does not refer to a particular system.  For example, the unitary quantum evolution on a Hilbert space ${\cal H}$ preserves the norm
   of
    vectors $|\psi\rangle\in {\cal H}$ and is invariant under global phase change. These facts could, but need not, be explicitly incorporated into the appropriate dynamical equations. In any case, the manifold of inequivalent initial conditions is actually the projective space $P{\cal H}$.

   This type of constraints should be clearly distinguished from those that are imposed as specific to a particular system. The latter are in fact a part of the specification of the system under consideration. Such constraints need not, but usually are, incorporated into an appropriate reformulation of the dynamical equations.
  An example  of such constraints is the constraint of the constant length of a pendulum in a constant gravitational field, which, if taken explicitly into the account, reduces the dimensionality of the manifold of initial conditions from 6 to 2. However, as pointed out, these types of constraints are in fact a part of the definition of what is the system under consideration and are treated in the next section.

 {\bf Remark 1}
Evolution of a quantum system on an $N$-dimensional Hilbert space ${\cal H}^N$ is equivalent to a linear symplectic flow
 on a symplectic manifold $P{\cal H}^{N-1}$ \cite{Ashtekar, Brody, Marmo}. Any such linear system as $N-1$ integrals of motion and is completely integrable. This can also be seen as a consequence of the fact that the group of automorphisms of such quantum system is $U(N)$, so that any pair of pure states can be connected by some unitary transformation. Any Hamiltonian, i.e. a Hermitian operator can be diagonalized using some $U(N)$ transformation.

Obviously, the notion of FDF is rather formal, and does not correspond to the one commonly  used in Physics. The later is captured by consideration
 of the system's structure and is precisely expressed using the relevant dynamical algebra.

\subsection{Structural degrees of freedom}

The notion of structural degrees of freedom is inspired by intuitively justified counting of independent elementary motions in compound classical Hamiltonian systems of mechanical type. A 1D particle is naturally associated one degree of freedom. If there are $N$  1D particles with the energy $H_i(q_i,p_i)=p_i^2/2m_i$ each, then a collection of such systems has the energy $H=\sum_i^N p_i^2/2m_i+V_{int}(q_1,q_2\dots)$, where the interaction is described by a function depending on some of the coordinates. Independently of the actual dynamics, the number of degrees of freedom
 of such a system is taken to be $N$. A quantum system obtained by quantization of such classical system is assumed to have the same number of
  degrees of freedom as the classical one. Let us stress that this notion of the number of degrees of freedom is not related to the inequivalent initial conditions for the dynamical equations, nor is related to the actual motion of the system. Rather, it is related to independent possible motions the system.
    A straightforward  definition of the number of the structural degrees of freedom for the systems with energy of the above form might be given as the number of momenta that appear quadratically in the energy. However, many realistic models of quantum as well as classical systems have the energy expressed in terms of natural dynamical variables which is not of the above form.   Examples of quantum systems are: the systems of spins, identical bosons or fermions etc..., and examples of classical systems are different ridged bodies and degenerate completely integrable systems written in the action-angle variables. Obviously, a more general definition of structural degrees of freedom is needed in order to apply the elementary intuition onto more complex systems.

{\it Dynamical algebra of relevant dynamical variables}

 A particular physical system is specified, and thus distinguished from
 an abstract general framework (such as "Hamiltonian dynamical systems", or "Unitary evolution"), by describing what can be measured on it, i.e. by specifying the set of basic dynamical variables, and by expressing the interactions within the system in terms of these dynamical variables.
In other words, the class of physically relevant dynamical variables
should be described and the evolution equations (the Hamiltonian) should be expressed in terms
of these dynamical variables.

Structure of the physical system is mathematically described by relations between the basic dynamical variables which form a Lie algebra $g$. Each of the
 basic dynamical variables is associated with an elementary change of the system's state, and successive independent changes are related by the Lie algebra
  structure of the set of basic dynamical variables.
 In classical mechanics the Lie algebra is realized by functions on a symplectic manifold and the Poisson bracket, and in  quantum mechanics by operators
 on a Hilbert space and the commutator. In either case, by  definition of what the dynamical algebra is, the manifold of the system's states is such that the action of the corresponding Lie group $G$ generated by the Lie algebra $g$, is transitive. The Lie algebra defined in this way is traditionally called the systems dynamical algebra. In our context, more appropriate name would be the structural algebra, but we shall use the traditional terminology.

 Description of a dynamical system amounts first to the specification of its dynamical algebra g and its manifold of states (providing an irreducible realization of $G$ as a  group of transformations). Once  this is done, the generator of the evolution, i.e.
 the Hamiltonian, is specified as an expression (possibly nonlinear) in terms of the basic dynamical variables belonging to g. Thus, in general
  the Hamiltonian is not necessarily an element of $q$.

In what follows the dynamical algebra $g$ will always be a  Lie
algebra, with rank $l$ and dimension $n$. The Lie group generated by the dynamical algebra will be denoted by $G$.

{\it Definition of the structural degrees of freedom }

The following considerations restricted onto the quantum systems with unitary irreducible representation of the dynamical algebra have been provided in \cite{Cheng1,Cheng2,Cheng3}. We present the definition of SDF first on the purely algebraic level, with no specification of the system quantum or classical nature. We than indicate how the classical and quantum systems with a given dynamical algebra are constructed.

Consider a system (classical or quantum) with a given dynamical algebra $g$ of distinguished variables. The notion of SDF is defined
in the same way for classical and for quantum systems, and is based only on the properties of the
 dynamical algebra.
 The
dynamical algebra $g$ has $\gamma$ different chains of
subalgebras: $g\supset g^l_{s^l}\supset g^{l}_{s^l-1}\dots \supset
g^{l}_{1},\>l=1,2,\dots \gamma$.
 Casimir elements
 of $g$ and all the algebras in (any of) the subalgebra chain form the relevant complete set of dynamical variables (CSDV) $Q_j, j=1,2\dots d$.
 There is $d=l+(n-l)/2$ of these,
 independently of the subalgebra chain.
Some of these Casimir elements are fully degenerate.

{\it Definition of SDF} The number of non-fully degenerate  elements in CSDV is $m\leq
(n-l)/2$ and is by definition the number of SDF. The non-fully degenerate Casimir elements in a particular
chain are the dynamical variables that define  $m$ SDF.
 The number of SDF is chain independent, although the elements that define  SDF are chain dependent.

The previous definition of SDF is given in algebraic terms with no reference to the classical or quantum nature of the system.
It remains to construct classical and quantum systems corresponding to the given dynamical algebra g, which would thus have the corresponding
 SDF. This is done using the KKS theory of orbits of the co-adjoined action of g \cite{KKS}. Given a dynamical Lie algebra g (from a certain class) one can show that orbits $O_g$ of the co-adjoined action of g are symplectic manifolds, which can be used as phase space ${\cal M}\equiv O_g$ of Hamiltonian dynamical systems with
   g as the dynamical algebra, and the corresponding SDF. To construct the quantum system with g as the dynamical algebra one can directly use the theory of unitary irreducible representations of g, or quantize the above symplectic manifolds $O_g$.
 Once the system phase space or the system Hilbert space are constructed, the system is finally specified by giving a Hamiltonian as an expression, possibly nonlinear, in
 terms of representatives of the elements of g. Observe that SDF are specified
  with no reference to a particular Hamiltonian, so that SDF are not related to the properties of the paricular system. Furthermore, ordering ambiguity, which occurs in the quantization of Hamiltonians, effects in no way the number and the type of SDF.

 {\bf Remark} In some particular realizations of an algebra the non-fully degenerate Casimir elements might
  not be independent. The physical system corresponding to this realization has the number of SDF  smaller than it is for the abstract algebra (Please see section 5.1.4).


{\bf Remark: Composite quantum systems} If the dynamical algebra $g$ of a quantum system $C$ can be represented as a direct sum of dynamical algebras of two systems $A$ and $B$,
 that is $g^C= g^A\oplus g^B$, then the tensor product of irreps of $G^A$ and $G^B$ is an irrep space of $G^C$, that is ${\cal H}^C={\cal H}^A\otimes {\cal H}^B$.
If $l_{A,B}$ and $n_{A,B}$ are the ranks and dimensions of  $g^A$
and $g^B$, then in general the number of SDF of $C$ is
$m_C=m_A+m_B$. Thus, in the case $g^C= g^A\oplus g^B$ the system
$C$ can be represented as a union of two  systems and the number
of SDF is additive. Of course, there are systems whose dynamical algebra is not decomposable in the above form, but which, nevertheless, have more
 than one SDF (Please see section 5.1.4).
Quite in general,  if the dynamical algebra $g$ of the system is
semi-simple then it can be uniquely expressed as a direct sum of
mutually commutative
 and orthogonal simple algebras: $g=\oplus_k g_k$ and
the Hilbert space which is an irrep space of $g$ factors as
$H=\otimes_k H_k$. Thus, in the case of semi-simple dynamical
algebra the number of SDF
 is additive, but the number of SDF in all the factor systems with $g_k$ dynamical algebras need not be unity for each $g_k$. Analogous statements
  apply to classical systems with semi-simple Lie dynamical algebra.

  {\bf Examples} Following examples will also be used later. The first example is relevant for classical as well as quantum systems.
   The second example is commonly associated with genuinely quantum systems, but nevertheless has perfectly well defined realization as a
    (classical ) Hamiltonian dynamical system.

  2) A system with Heisenberg-Weyl $h_4$ dynamical algebra: The algebra
   is commonly given by the relation between its four basis elements $\{a^{\dag},a,n,I\}$
\begin{eqnarray}
   &&[a,a^{\dag}] = I,\quad [n,a^{\dag}]=a^{\dag},\quad [n,a]=-a,\nonumber\\
   &&[a^{\dag},I] = 0,\quad [a,I]=0,\quad [n,I]=0.
\end{eqnarray}

$h_4$ has $rank=2$, and is not semisimple algebra. The algebra chain $h_4\supset u(1)\otimes u(1)$ determines the SDF.
The Casimir elements of $h_4$ and $u(1)$ are proportional to unity, so there is only one SDF. In the classical case the phase space is $R^2$ with symplectic coordinates $(q,p)$. The Casimir element corresponding to one SDF is in fact the action variable of the harmonic oscillator $J=p^2+q^2$.
  In the quantum case the algebra $h_4$ is uniquely represented by multiplication and differentiation operators on $L_2({\bf R})$, which are related to the basis elements $a,a^{\dag}$
   by $ a=(x+i\partial/\partial x)/\sqrt {2\hbar}, \>a^{\dag}=(x-i\partial/\partial x)\sqrt {2\hbar}$.  The Casimir element corresponding to the one SDF is the number operator $\hat n$.

 A classical mechanical systems, composed of two $h_4$ subsystems, has  the dynamical algebra $h^1_{4}\otimes h^2_{4}$ and the canonical variables $p_{x_1},p_{x_2},x_1,x_2$ satisfying
 \begin{equation}
 \{p_{x_i},x_j\}=\delta_{ij},\> \{x_i,x_j\}=0,\> \{p_{x_i},p_{x_j}\}=0,\quad i,j=1,2,
 \end{equation}
 where $\{,\}$ denotes the Poisson bracket, and the indexes $i,j=1,2$ denote the first and the second system.
 The basic canonical variables are related to the algebra basis elements by $a_j=(x_j+ip_{x_j})/\sqrt {2}, \> a_j^{\dag}=(x_j-ip_{x_j})\sqrt {2},\>j=1,2,\>i=\sqrt {-1}$. The system has two SDF corresponding to the subalgebra $h^1_4\otimes h^2_4=u^1(1)\otimes u^1(1)\otimes u^2(1)\otimes u^2(1)$. The same applies
  to the corresponding quantum version.

  Hamiltonian of two possibly interacting oscillators is expressed in terms of $p_{x_i},x_i,\>i=1,2$ as $H=p_{x_1}^2/2m_{1}+p_{x_2}^2/2m_2+V(x_1,x_2)$.
In the case that there is no interaction $V=0$ the Hamiltonian is, in the classical as well as the quantum case, a linear expression of the algebra generators $H\sim\sum_{i=1,2}\omega_i n_i$. The variables corresponding to the chosen SDF, i.e. the algebra generators $n_i,\>=1,2$ are constants of
 motion for such Hamiltonian dynamics. In other words, the subalgebra that is used to define the SDF is in fact a dynamical symmetry of the system.
 This dynamical symmetry is usually  broken if there is interaction between the variables corresponding to different SDF. However,
  if the interaction between the harmonic oscillators is of the form $V(x_1,x_2,p_{x_1},p_{x_2})=(x^2+p_{x_1}^2)(x_2^2+p_{x_2}^2)=n_1n_2$ then the subgroup generated by the subalgebra used to define the SDF is a dynamical symmetry.

  2) A system with $su(2)$ dynamical algebra: The algebra is generated by the Pauli spin matrices $\{I,\sigma_1,\sigma_2,\sigma_3\}$ that satisfy the relations
  \begin{equation}
  [\sigma_k,\sigma_l]=2i\sum_m\epsilon_{k,l,m}\sigma_m,\quad [\sigma_{l},I]=0\quad k,l,m=1,2,3.
  \end{equation}
 The dimension and the rank of $su(2)$ are $d=3,l=1$, so that the number of SDF of a system with this dynamical algebra is one. The subalgebra chain
  $su(2)\supset u(1)$ determines the SDF, given by the non-degenerate Casimir $\sigma_3$.

  A pair of spins  has the dynamical algebra $su^1(2)\oplus su^2(2)$. The number od SDF is two. The SDF could correspond to the subalgebra chain
  $su^1(2)\oplus su^2(2)\supset u^1(1)\oplus u^2(1)$ with non-degenerate Casimirs $\sigma^1_3$ and $\sigma^2_3$.
 Wether the subalgebra corresponding to the two SDF  generates a dynamical symmetry depends on the Hamiltonian and the particular  form of the interaction between the two spins. This is the case if the interaction involves only the subalgebra generators $\sigma^1_3$ and $\sigma^2_3$ of the form $\sigma^1_3\otimes \sigma^2_3$ (Please see section 5.1.2).

\subsection{ Dynamical degrees of freedom}

 The notion of dynamical degrees of freedom is central in our work. Although the formal definition is quite simple, actual determination of
  the number of DDF, without previous solution of the dynamical problem, seams to be impossible. However, it is worth giving a precise definition and attempt
   to use it in characterizing the relation between the properties of quantum and classical dynamics.
   We have seen that the Hamiltonian need not be an
   element of the dynamical algebra i.e. might be given by a nonlinear expression in terms of the basic dynamical variables. In this case, the coadjoined orbits $O_g$ of the dynamical algebra $g$, discussed after the definition of SDF, need not be invariant under under the Hamiltonian dynamics.
   Manifolds which are dynamically invariant and irreducible might be of dimension larger than  the number $N$ of SDF.
  For example, if the Hamilton's function is the only constant of motion then a generic irreducible invariant manifold is $2N-1$ dimensional manifold of constant energy.
   Similarly, an irreducible invariant manifold of a composite quantum system might be of larger dimension than the total number of additive
  SDF. This happens for example if the interaction generates entanglement between the components.
 It makes sense to analyze, as an important dynamical property of the systems, the dimension of its generic invariant manifold.

 {\it Definition of DDF} The number of DDF of a quantum or classical system with the SDF defined as above is by definition the dimension of its generic dynamically invariant and irreducible manifold.

 In classical mechanics the generic dynamically invariant and irreducible manifold is understood as the manifold in which a generic orbit is embedded,
  so that the number of SDF is in fact the embedding dimension of a generic orbit. In quantum mechanics this is the manifold (subspace) on which the generic orbit is ergodic.  The difference occurs because of the nonlinear character of the classical dynamical equations and the consequent fact that the generic orbit of the classical system might be a (fat) fractal \cite{Farmer, usIAN}. This is impossible in quantum mechanics where all orbits of the state vector are either periodic, with different dimensionality, or quasi-periodic, i.e. ergodic on a torus of dimension equal to the half of FDF. Of course, from a practical point of view a periodic orbit with very long period is for some purposes like an ergodic one. However, for our purposes the ergodicity of quantum orbits is a qualitative property meant to characterize the orbits with generic behavior.

 Consider an integrable Hamiltonian system with $N$ SDF. In the generic case the number of DDF is also $N$. However, the system might be degenerate in such a way that all frequencies of motion on all invariant dim-$N$ tori are commensurate so that all orbits (except few isolated ones) are periodic
  and therefore the system has only one  DDF. Analogously,
 consider a quantum system with an $N$-dimensional Hilbert space (as a special case of an integrable system). If the energy spectrum of the system is known then the number of DDF can be discussed without specifying the dynamical algebra and the number of SDF. If $N-1$ ratios $e_i/e_1$ of the energy eigenvalues $e_i, i=1,2\dots N$ are irrationally related then typical orbits (except the stationary ones) are quasi-periodic, and the motion of the state vector is ergodic on
  an $N-1$ dimensional torus in ${\bf R}^{2N}$.
   The number of DDF is $N-1$, the same as the number of (quantum) FDF. If there are some rational relations the orbit in the union of the corresponding eigenspaces is periodic, and the number of DDF is smaller than $N-1$. In the special case of $N=\infty$ and when all eigenvalues are multiples of a single value, the generic orbit is a circle of dimension $1$. This corresponds to the harmonic oscillator with the
    dynamical algebra $h_4$ and the number of SDF and DDF both equal to unity.
     The number of FDF is of course infinite.
    It appears, from number-theoretic reasons and linearity, that a generic quantum system with $N$ dimensional Hilbert     space should have $N-1$ DDF.

     Obviously, the number of SDF  is smaller or equal than the number of FDF, but the number of DDF can be smaller, equal or greater than the number of SDF. We have seen that the NSDF of a classical system and its quantization are equal. However, determination of the relation between NDDF of
related quantum and classical systems is a challenging problem. The problem involves analyzes of the relation between classical integrability, separability in the fixed set of SDF, dynamical symmetry and generation of quantum entanglement between the fixed set of SDF. In what follows we shall contribute very little to the general solution of this quite difficult problem. We shall only use a couple of examples to indicate some important facts which certainly contribute to the impression that there should be a general relation between
 NDDF of related quantum and classical systems, but also show that the relation is not a simple one. In order to proceed with the analyzes we shall
  need to define a) a measure of quantumness of a given quantum state and  b) a classical system which is naturally  associated with an arbitrary quantum system with defined dynamical algebra and the Hamiltonian operator. The well known notion of generalized coherent states is needed in both of these steps, so we shall
   provide a brief recapitulation.

 \section{ g-coherent states and the measure of quantumness}

 The number of SDF is in the quantum case directly related to the dimensionality of the manifold of the coherent states corresponding to the
  system's dynamical algebra. The coherent states corresponding to a dynamical algebra $g$ will be denoted as $g$-coherent states.
   Let us briefly recapitulate the definition of the generalized g-coherent states. This is also relevant for the
   definition of the classical model of a quantum system, to be introduced in the next section.

Consider first  quantum systems with $g=h_4\oplus h_4\dots\oplus h_4$ dynamical algebra and the Hilbert space
 ${\cal H}=L_2(x_1)\otimes L_2(x_2)\dots L_2(x_N)$. It makes sense to define a level of quantumness of a state  $|\psi\rangle\in {\cal H}^N$ by
  the expression
  \begin{equation}
  \Delta_{h_4}(\psi)=\sum_i^N \Delta_{\psi} \hat x_i\Delta_{\psi}\hat p_{i},
  \end{equation}
  where $\Delta_{\psi} \hat A$ denotes dispersion of $\hat A$ in $|\psi\rangle$.
  The states that have minimal quantumness $\Delta_{h_4}$ are $N$-products of Glauberg $h_4$-coherent states.  This is one of the important properties of the $h_4$ coherent states $|\alpha_1\rangle$ carried over to the products of $h_4$-coherent states $|\alpha_1\rangle\alpha_2\rangle\dots|\alpha_N\rangle$. In the previous expressions $\alpha_i\in {\mathbf C}$ are complex parameters that parameterize the manifold of the coherent states for the $i$-th system.  It is well known that all such states can be obtained by the
   action of a displacement operators $D(\alpha_1,\dots \alpha_N)=\exp \sum_i^N(\alpha_i\hat a_i^{\dag}-h.c.)$ onto the vacuum state.
 It is also well known that  this property is used to define and construct generalized $g$-coherent states for systems with compact dynamical
  groups with finite ${\cal H}^N$.  However, in this case the level of quantumness $\Delta_{g}(\psi)$ is defined in a different way.

  Let us first, very briefly, recapitulate the construction of generalized $g$-coherent states.
   There are several  generalizations of Glauber, i.e. $h_4$ coherent states.
 Perelomov \cite{Perelomov}
 and Gillmore \cite{Gilmor} independently introduced two different generalizations based on the group-theoretical structure of the $H_4$ coherent states.
  The essential
 ideas of both approaches are the same, the differences being in the class of Lie groups, and the corresponding available tools, and in
  the choice a reference state.
In both approaches, the set of $g$-coherent states depends,
besides the algebra $g$, also on the particular Hilbert space
$H^{\Lambda}$ caring the irrep $\Lambda$ of $g$ and on the choice
 of an, in principle(Perelomov), arbitrary referencee state, denoted $|\psi_0>$.
  Here $\Lambda$ is a multi-index   indexing irreps of $g$.

     The subgroup $S_{\psi_0}$ of $G$ which
  leaves the ray corresponding to the state $|\psi_0>$ invariant is called the stability subgroup of
 $|\psi_0>$: $h|\psi_0>=|\psi_0>\exp i\chi(h), h\in S_{\psi_0}.$
 Then, for every $g\in G$ there is a unique
 decomposition into the product of two elements, one from $S_{\psi_0}$ and one from the coset $G/S_{\psi_0}$ so that
  $g|\psi_0>= \Omega|\psi_0> \exp i\chi(h)$.
 The states of the form $|\Lambda, \Omega>=\Omega|\psi_0>$ for all $g\in G$ are the $g$ coherent states. The notation indicates the fixed arbitrary irrep by $\Lambda$ and the particular element from the coset $G/S_{\psi_0}$ by $\Omega$. The latter parameterizes the set of $g$-coherent states, obtained using the $\Lambda$ irrep of $g$.
  Geometrically the set of $g$-coherent state
 form a manifold with well defined Riemanien and symplectic structure.

Consider a quantum system with the dim-n dynamical algebra represented on a finite-dimensional Hilbert space ${\cal H}^N$.
Denote by $L_i,\> i=1,2,\dots n$ the algebra generators in an arbitrary basis of the algebra.
Level of quantumness of a pure state $|\psi>\in {\cal H}^N$  is defined as
\begin{equation}
\Delta_g(\psi)=\sum_i^n<\psi|L_i^2|\psi>-<\psi|L_i|\psi>^2,
\end{equation}
where the sum is taken over an orthonormal bases of the dynamical
algebra $g$. It make sense to consider the quantity $\Delta_g(\psi)$
as a measure of  quantumness of the state $\psi$ for the system with the dynamical algebra $g$. The general definition of $g$-coherent states is such that $\Delta_g(\psi)$ is
 minimized precisely by such coherent states.

 In the case of systems with a semi-simple dynamical algebra $g$,  studied by Gillmore,
  the irrep space is characterized by the unique
 highest weight state $|\Lambda,\Lambda>$ (or the lowest weight state $|\Lambda,-\Lambda>$).
  This vector is annihilated  all $E_{\alpha},\> ((E_{-\alpha})$  where $E_{\alpha}$ ( $E_{-\alpha}$) belong to the standard Cartan basis of $g$: $\{H_i,E_{\alpha},E_{-\alpha}\}$..
  The state  $|\Lambda,\Lambda>$ is left invariant by  operators in the Cartan subalgebra $H_i$.    The set of $g$ coherent states
 can be represented in the form of an action of the displacement operator
 on the reference state $|\Lambda,\Lambda>$.
\begin{equation}
|\alpha>=D(\alpha)|\Lambda,\Lambda>=\exp[\sum
\alpha_iE_i-h.c.]|\Lambda,\Lambda>,
\end{equation}
where$\alpha$ is a multi-parameter standing for the set of complex parameters $\alpha_i$ and the sum extends over
all $E_{-\alpha}$ that do not annihilate $|\Lambda,\Lambda>$. The
stabilizer $S_{\psi_0}$ of the reference state $|0>$ is the
subgroup
 generated by the Cartan subalgebra of $g$
The complex parameters $\alpha_i,i=1,2\dots M$ parameterize $2M$
dimensional manifold $G/S_{\psi_0}$ of $g$-coherent states.

It is obvious that the number of SDF of a quantum system is equal to half the number of dimensions of the manifold of g-coherent states.
However, if the Hamiltonian is a nonlinear expression in terms of the dynamical algebra generators then the manifold of g-coherent states is not
   dynamically invariant. Therefore the system that has started from  the manifold of $g$-coherent state,
     i.e. the states with minimal quantumness will necessarily leave this manifold and the quantumness will increase.
     Of course, later evolution, i.e. from non-coherent states, might lead to local (in time) increase or decrease of quantumness. In any case,
    $g$-quantumness  is not  preserved along typical orbits. Quantum systems that generate g-quantumness, i.e. such that $g$-quantumness is not a constant of motion,
  satisfy the inequality NDDF$>$NSDF.  We see again that the number of DDF is often larger than the number of SDF.

\section{ Classical model of a quantum system}

Consider a quantum system given by a dynamical algebra $g$, a Hilbert space ${\cal H}$ and a Hamiltonian $\hat H$.
Classical Hamiltonian dynamical system on the symplectic manifold of g-coherent states
$G/S_{|\psi_0>}$ and given by the Hamiltonian function $H(\alpha)=<\alpha|\hat H|\alpha>$ is called the classical model
of the quantum dynamical system $({\cal H},g,{\hat H})$. Physically, the classical model corresponds to a certain type of coarse-grained description of the quantum system. The coarse-graining is modeled by imposing specific constraints on the Hamiltonian formulation of quantum dynamics \cite{usPRA1,usPRA2}.

In the Hamiltonian formulation of a quantum system  $({\cal H}^N, g,\hat H)$ \cite{Ashtekar, Brody, Marmo}, the Schr\" odinger equation on ${\cal H}^N$ (the dimension of (${\cal N}^N$, can be $N=\infty$) is considered as a Hamiltonian dynamical system on the manifold ${\cal M}={\mathbf R}^{2N}$, with the simplectic and Riemannian  structures given by the imaginary and the real part of the Hermitien scalar product. The Hamiltonian of the system is given as $H(X)=\langle \psi_X|\hat H|\psi_X\rangle$. The set of real and imaginary components $\Re c_i,\> \Im c_i$ of a state $|\psi\rangle$ in any basis gives a set of canonical coordinates $Q_i=\Re c_i/\sqrt 2$ and
 $P_i= \Im c_i/\sqrt 2$ of the point $X_{\psi}$.  The dynamical algebra of the quantum system is $g$ with the generators $L_i$ and the corresponding $g$-coherent states. Consider a constrained quantum Hamiltonian system with the constraint that the quantumness $\Delta_g(\psi)$ is preserved minimal during the evolution, and call such system $\Delta_g$ constrained. The manifold of constraints $\Delta_g(\psi)=min$ is denoted $\Gamma_g$.

{\it Definition of the classical model}  The classical model of the quantum system $({\cal H}^N, g,\hat H)$ is by definition the reduction of the $\Delta_g$ constrained Hamiltonian system
  $({\cal M}, \langle\hat H\rangle)$  onto the constrained manifold $\Gamma_g$.

 The constraint $\Delta_g$ depends on the algebra of basic dynamical variables and is related to an equivalence relation between the quantum pure states.
 For a given algebra the space of quantum pure states can be partitioned into equivalence classes such that  each equivalence class contains one and only one state from the constrained manifold, i.e. a state with minimal g-quantumness. The $\Delta_g$ constrained Hamiltonian system preserves the
 equivalence of states during the evolution.

One can use the theory of constrained Hamiltonian systems developed by Dirac, to study the Hamiltonian system with $\Gamma_g$ constraints. However,
 if $\Gamma_g$ is symplectic, the general procedure can be bypassed, with the known result that the reduced constrained system is in fact also Hamiltonian on $\Gamma_g$ with the Hamilton's function given by $\langle \hat H\rangle|_{\Gamma_g}$. In our case, when $\Gamma_g$ is determined
  by $\Delta_g$, it is known that $\Gamma_g$ coincides with the manifold of $g$-coherent states and the latter is a symplectic manifold.
  Thus, a definition of the classical model equivalent to the previous one can be given as

  {\it Definition of the  classical model} The classical model of the quantum system $({\cal H}^N, g,\hat H)$ is the Hamiltonian dynamical system on the symplectic manifold
   parameterizing the $g$ coherent states $\alpha\in \Gamma_g$  with the Hamilton's function $H(\alpha)=\langle\alpha|\hat H|\alpha\rangle$.

 The classical model has the same dynamical algebra and the same number of SDF as the corresponding quantum system.

  Special examples of the construction of classical models for a system of coupled unharmonic oscillators and spin-$j$ systems with arbitrary $j$
    that is the systems with dynamical algebras $h_4\oplus h_4\oplus\dots h_4$ and $su(2)$ on ${\cal H}^{2j+1}$, have been introduced in
  \cite{usPRA1} and \cite{usPRA2} respectively.  In the following we shall treat in detail the example of two coupled oscillators, and consider two more examples given by $su(2)\oplus su(2)$ and $su(3)$ dynamical algebras.

  Classical limit of a quantum system, if i exists,  is obtained from the corresponding classical model in some macro-limit. Given a quantum system, its classical model and its classical limit are not the same Hamiltonian dynamical systems. For the oscillators systems the macro-limit corresponds to $\hbar\rightarrow 0$ and for the spin system to $j\rightarrow \infty$. The macro-limit is well controlled because the classical model satisfies the constraint of constant and minimal dispersions.    Dynamics of the classical model and the classical limit might be qualitatively different.

  {\bf Remark} Structural features of systems of identical bosons or fermions are also described by appropriate dynamical
   algebras. Consequently, the definition of SDF, construction of the corresponding g-coherent states and finally the
construction of the corresponding classical model are all performed as in the general case. Many different dynamical algebras have been constructed using the canonical commutation relations (CCR) for a system of bosons or fermions
  \begin{equation}
  [a_i,a_j^{\dag}]_{\pm}=\delta_{ij},\>[a_i^{\dag},a_j^{\dag}]_{\pm}=0,\>[a_i,a_j]_{\pm}=0,
  \end{equation}
   and have been found useful in the treatment of dynamical problems. In (7) $a_i^{\dag}$ is the single particle creation operator of the $i$-th energy level, and + (-) indicates the anti-commutator (commutator). Generators of the relevant dynamical algebras are expressed in terms of single particle operators and their bilinear combinations. Most  commonly used  dynamical algebras, describing the structure of systems of $r$-level fermions or bosons,  are $u(r)$ or $so(2r)$ and $(sp(2r)$ modeling physically different situations. For example $u(r)$ generators (not all independent) are given by the set  $\{a_i^{\dag}a_j,\>i,j=1,2\dots r\}$, with the corresponding commutation relations. We shall provide few more details only in the $u(r)$ case.

    Subalgebra structure, which determines the number of SDF, of the representations of these groups can be very  reach, and strongly depends on representation. An example is given by the case when the
    elementary excitation operators for a system of $n$ $r$-level fermions are of the form
    \begin{equation}
    \{a_i^{\dag}a_k,\> 1\leq k\leq n, n+1\leq i\leq r\}.
    \end{equation}
    and these represent $n(r-n)$ SDF of such fermion system with $u(r)$ dynamical algebra.

 Typical Hamiltonian of a many-body system of identical particles is expressed in terms of the algebra generators, by possibly nonlinear expression
   \begin{equation}
   H=\sum^r_{i}\omega_i a_i^{\dag} a_i+\sum^r V_{ijkl}a_i^{\dag} a_j^{\dag} a_k a_l.
   \end{equation}
   Ground state of such a Hamiltonian is denoted by $\psi_0$ and of course shares the symmetry of the system. This fact introduces crucial difference is the manifolds of coherent states for bosons or fermions with the same $u(r)$ dynamical algebra.

   Generalized coherent states are given by the the action of the exponential mapping from the set of elementary excitations on the ground state
   \begin{equation}
   |\alpha\rangle=D(\alpha)|\psi_0\rangle=\exp\sum_{ik} (\alpha_{ik}a_i^{\dag}a_k+\bar \alpha_{ik}a_k^{\dag}a_i)\>|\psi_0\rangle,
\end{equation}
  and, in the fermionic case, form the symplectic manifold $U(r)/U(n)\otimes U(r-n)$, which is the phase space of the classical model of the fermionic system. On the other hand the manifold of the coherent states, i.e. the phase space of the classical model in the bosonic case is $U(r)/U(1)\otimes U(r-1)$. Classical model share the symmetry of the ground state, that is the symmetry of the fermionic or bosonic system. In any case the coherent states are considered as the most classical states of the considered quantum system.
   These states minimize the quantumnes $\Delta_{u(r)}(\psi)$ defined as in the general case (5).
   Well defined classical model for fermions or bosons does not imply existence of an appropriate classical limit.
   The later is achieved as the number of particles $n\rightarrow \infty$ only for bosons.
   Of course the set of coherent states is not invariant on the quantum evolution generated by the Hamiltonian with general $V_{ijkl}\neq 0$, and the quantumness is not preserved by such evolution.
    Dynamics of the classical model is given by the general prescription. The Hamilton function on the phase space is the expectation $\langle \alpha|\hat H|\alpha\rangle$. Dynamics of examples of such classical Hamiltonian systems corresponding to bosons or fermions have been treated in \cite{Cheng3}. Relation between the evolution of quantumness and the qualitative properties of the classical model has not been studied.

\section {Relation between DDF of a quantum system and its classical model}

 Dynamics of classical
models of quantum systems have been studied for
various examples in \cite{Cheng1,Cheng2,Cheng3}, but without understanding them as constrained Hamiltonian systems. Relation between
dynamics of entanglement and the dynamics of classical models for
a pair of qubits was studied in
\cite{JaPhysRev06, JaAnnPhys}. Here we want to argue in favor of existence of a general relation between the generation of g-quantumness and
 qualitative properties of the dynamics of the classical model as defined in the previous section, and to use this to infer relations between
  the corresponding NDDF.

\subsection{Examples}

Our first example is rather trivial in that it illustrates the case when the quantum system and its classical model are the same Hamiltonian dynamical system. The second example considers a pair of $1/2$-spins as a genuinely quantum system, i.e. a quantum system which can not be obtained by quantization of a classical mechanical system. Nevertheless the classical model is well defined and the question of relation between NDDF for the two systems makes sense. Furthermore, in this case the question has a definite and complete answer,  suggesting a general relation between the integrability of the classical model, the dynamical symmetry and the dynamical generation of quantumness. In this case, the NDDF of the spins system and the classical model are equal for any of the different dynamical regimes. However, this simple relation between integrability of the classical model and the lack of quantumness generation is not valid in general. The relation is more complicated as is illustrated by our third example of coupled nonlinear oscillators.

\subsubsection{ von-Neumann case: $u(N)$ dynamical algebra}

This is a rather trivial example because the quantum system and its classical model are identical. The classical model is integrable, the quantum Hamiltonian is an element of $u(N)$  dynamical algebra  and
 all states of the quantum system have equal $u(N)$-quantumness.

The quantum system is described by $N$ dimensional Hilbert space ${\cal H}^N$ and the dynamical algebra $u(N)$, which means that every hermitian operator
 on  $H^N$ has physical interpretation as a measurable quantity. In particular, any Hamiltonian is an element of the dynamical algebra, i.e. a Hermitian operator.
  Due to the normalization and global phase invariance the state space of the system
 is $CP^{N-1}$ which is topologically like $S^{2N-1}/S^1$, and represents a $2(N-1)$ manifold with Riemanien and symplectic structure. Geometrically,
 it should be natural to associate $N-1$ FDF with this system. The same number of SDF follows from $u(N)$ dynamical algebra. The Hilbert space
 is the fully symmetric irrep space of $u(N)$ with the highest weight: $\Lambda=(1,0,\dots 0)$. The basis can be labeled by the following chain
 of subalgebras: $u(N)\supset u(N-1)\dots\supset u(1)$ with the corresponding  Casimir elements $C_i^{u(k)},\> i=1,2\dots k, \>k=1,2\dots N$. The
$N-1$ non-fully degenerate operators are $C_i^{u(k)},\> i=1,2\dots
k, \>k=1,2\dots N-1$ and label the basis $|i\rangle=|0,0,\dots i,\dots
0\rangle,i=0,1,2,\dots N-1$.
Explicitly: $C_k^{u(k)}|i>=\Theta(k-(N-i))|i\rangle$, and $\Theta(i)$ is
the Heaviside function on $i=1,2\dots N-1$. Thus there is $N-1$
SDF, the same as the number of
 FDF.

 Elementary excitation operators are given by:
$E_{i0}|\psi_0\rangle=|i\rangle,i=1,2,\dots N-1$ where $|\psi_0\rangle$ is the
lowest weight vector of the $\Lambda=(1,0,\dots 0)$
representation, and $u(N)$ coherent states are obtained as
$|\alpha\rangle=exp(\sum \alpha_i E_{i1}-h.c)|0\rangle$. Coherent states are
parameterized by the coset space $U(N)/U(N-1)\otimes U(1)$ which
is isomorphic to $CP^{N-1}$. We see that all
 states are $u(N)$ coherent states and therefore all states have the same quantumness.
   Any Hamiltonian can be diagonalized by an $U(N)$ transformation and thus expressed as a combination of the Casimir elements.

  It should be noticed that since any state is $u(N)$ coherent
 state the dynamics of the quantum system on $CP^{N-1}$ with the Hmiltonian $\hat H$, and its classical
 model with the Hamiltonian function $\langle\psi|\hat H|\psi\rangle$ on the phase
 space $U(N)/U(N-1)\otimes U(1)\sim CP^{N-1}$ are identical (and integrable)
 for any Hamiltonian.
A quantum system with generic $u(N)$ Hamiltonian and the corresponding (isomorphic) classical model have equal number of DDF.

\subsubsection{ Coupled spins: $g=su^1(2)\oplus su^2(2)$ semi-simple dynamical
algebra}

Consider a pair of spins with the Hilbert space $H=C^{2}\otimes
C^{2}$ and the Hamiltonian
\begin{equation}
 H=\omega(J_z^1+J_z^2)+\mu_x J_x^1J_x^2+\mu_z J_z^1J_z^2.
\end{equation}
where $J^{i}_{x,y,z},\>i=1,2$ are operators satisfying  the commutation relations of $su^1(2)\oplus su^2(2)$.

 The dynamical group of the system is $SU^1(2)\otimes SU^2(2)$. A possible subgroup chain is
  $SU^1(2)\otimes SU^2(2)\supset SO(2)\otimes SO(2)$. The number and the nature of SDF are as in the $su^1(2)\oplus su^2(2)$ example of section 2.2.
   The set of g-coherent states is given by products of su(2)-coherent states for each of the spins. Topologically,
 they represents a Cartesian product of two 2D-spheres $S^1\times S^2$ where the superscripts correspond to the spins 1 and 2.
 The set of coherent states is parameterized by local symplectic coordinates on  $S^1\times S^2$ as $|q,p\rangle\equiv |q_1,p_1,q_2,p_2\rangle$.
 They minimize quantumness (5). Non-coherent states are
  entangled with respect to SDF given by $J_z^1$ and $J_z^2$,  with the quantumness larger than minimal. The radius of each of the spheres of the coherent states is $2\sqrt{J\hbar}$ where $2J+1$ is the dimension of the relevant $su(2)$ representation.

 When $\mu_x=0,\mu_z\neq 0$ the Hamiltonian commutes with $J_z^1$ and $J_z^2$, and the system has the dynamical symmetry of $SO^1(2)\otimes SO^2(2)$ corresponding to the selected SDF. This is manifested in the fact that the system does not generate quantumness with respect to the considered
SDF despite the interaction $\mu_z J_z^1J_z^2$  between the two spins.

 If $\mu_x\neq 0$ the $SO^1(2)\otimes SO^2(2)$ dynamical symmetry is broken.
  The set of coherent states is
  not dynamically invariant. The system generates quantumness in the form of entanglement between the  $SO^1(2)\otimes SO^2(2)$
 dynamical degrees of freedom (please see fig. 1).
\begin{figure*}
\centering
\includegraphics[width=1\textwidth]{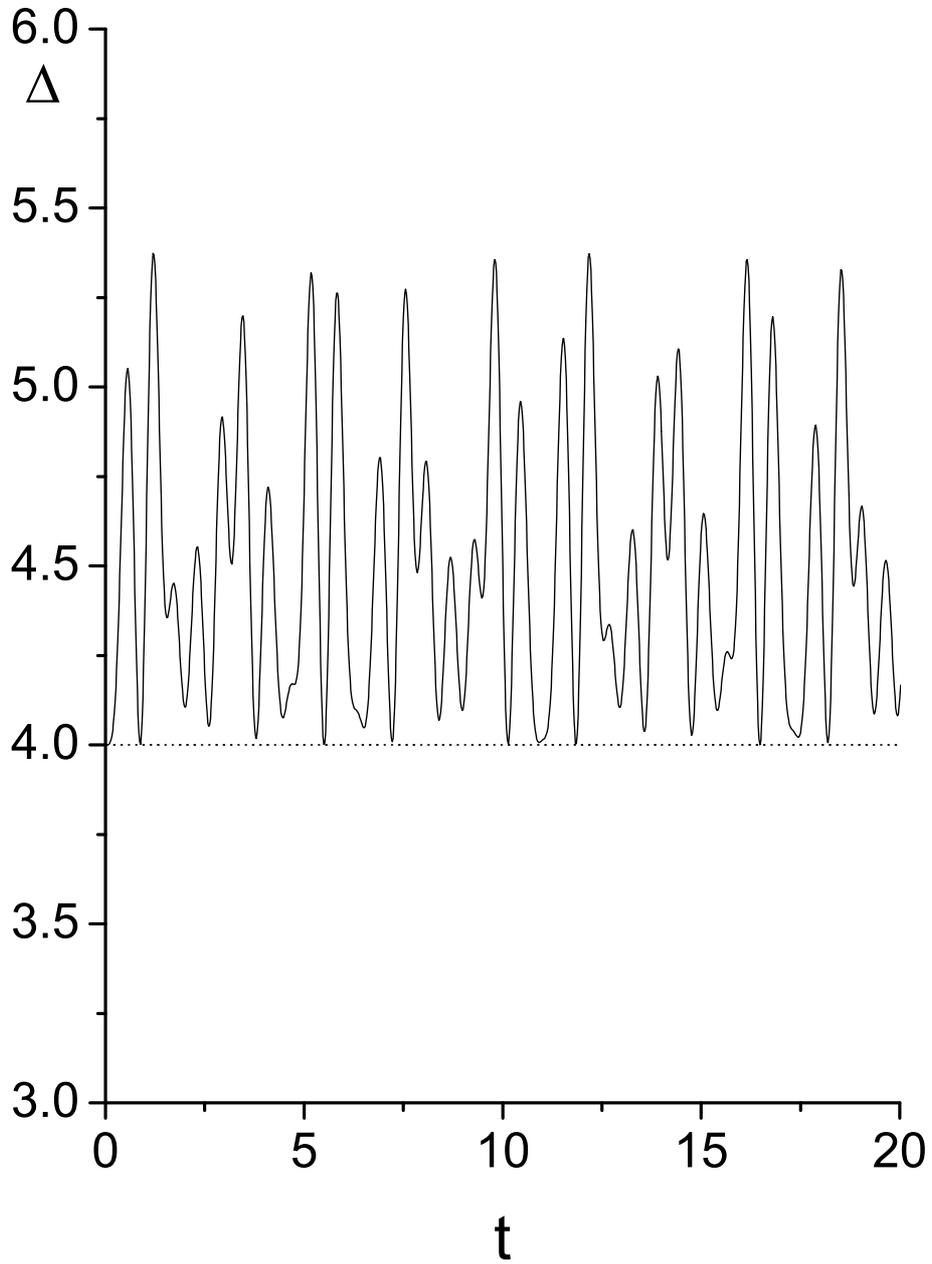}
\caption{ Illustrates dynamics of the quantumness
$\Delta_{su(2)\oplus su(2)}(t)$ with the Hamiltonian (11), starting from an
$SU(2)\otimes SU(2)$ coherent state with $J=1/2$.
Full line corresponds to  $\mu_x\neq 0$
and dotted to  $\mu_x=0,\mu_z\neq 0$.}
\end{figure*}

 {\it Dynamics of the classical model}

The phase space of the classical model is the manifold of $su^1(2)\oplus su^2(2)$ coherent states, i.e. the Cartesian product of two spheres. Local symplectic coordinates on $S^1\times S^2$ are denoted by $q_1,p_1,q_2,p_2$.
 The Hamilton's function of the classical model is the coherent state expectation $H(q,p)=\langle p,q|\hat H| p,q\rangle$ of (11), and is expressed in terms of
  the expectation  values of the generators
 $\langle p,q |J_{x,y,z}^{1,2}|p,q\rangle$ by the following formulas:
 \begin{eqnarray}
\langle\hat J^i_x\rangle(p,q)&=&{q_i\over 2}(4J-q_i^2-p_i^2)^{1/2},\nonumber\\
\langle\hat J^i_y\rangle(p,q)&=&-{p_i\over 2}(4J-q_i^2-p_i^2)^{1/2},\nonumber\\
\langle\hat J^i_z\rangle(p,q)&=&{1\over 2}(q_i^2+p_i^2-2J), \quad i=1,2
\end{eqnarray}
 where $J=1/2$ in our case of two $1/2$-spins.

  The Hamiltonian of the classical model   is given by
 \begin{eqnarray}
 H&=&(p_1^2+p_2^2+q_1^2+q_2^2)+\mu_z (p_1^2+q_1^2-2J)(p_2^2+q_2^2-2J)\nonumber\\
 &+&\mu_x q_1q_2[(4J-p_1^2-q_1^2)(4J-p_2^2-q_2^2)]^{1/2},
 \end{eqnarray}
 with $J=1/2$. Observe that the classical model involves only the coherent states expectations of $J^i_{x,y,z}$ and products of such expectations
  for different spins. No expectations of operators which are nonlinear expressions in terms of $J^i_{x,y,z}$ occur. It is often stated that the
   classical limit of the spin system is obtained by taking $J\rightarrow \infty$. Strictly speaking, this corresponds to the
   limit of classical models of a sequence of large quantum spins.
    The systems of 1/2-spins do not have the classical limit but do have classical model, i.e.  (13) with $J=1/2$.

 When $\mu_x=0,\mu_z\neq 0$, despite the interaction between the SDF, the Poisson bracket $\{H,\langle J_z^{1,2}\rangle\}$ is zero, being proportional to $\mu_x$. Like in the quantum case
  the system has dynamical $SO(2)\times SO(2)$ symmetry. The classical model is completely integrable with the obvious set of independent constants of motion.

  When $\mu_x\neq 0$ the $SO(2)\otimes SO(2)$ dynamical symmetry is broken, and the coherent state expectations $\langle J_z^{1,2}\rangle$ are not constants of motion any more.  The classical model might not be
   completely integrable.
   Chaotic orbits of the classical model with $J=1/2$ are easily found in numerical computations, and one such orbit is illustrated in fig.2.

   \begin{figure*}
\centering
\includegraphics[width=1\textwidth]{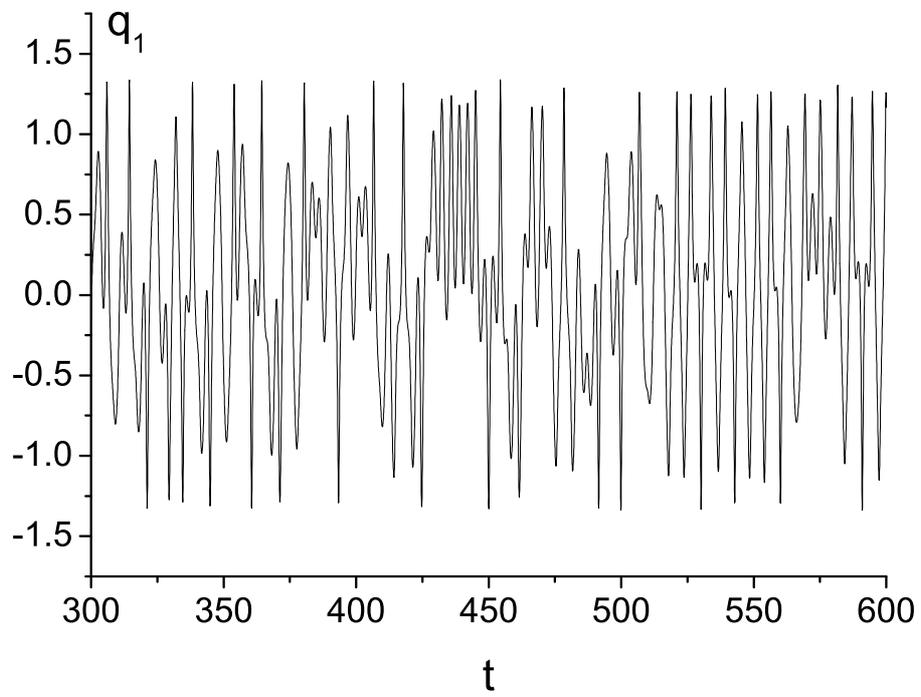}
\caption{ Illustrates $q_1(t)$ component of a chaotic orbit of the classical model with the Hamiltonian (13) for $J=1/2$ and $\mu_z=0,\mu_x=1$. }
\end{figure*}

   Thus, analyzes of the quantum system (11) and its classical model (13) suggest that the quantum system (11) generates quantumness if and only if the classical model (13) is not completely integrable. However, the following example  will show that such relation is not generally true.

    Nevertheless, one property of the quantum system and its classical model should be observed and stressed.  The two interacting $1/2$-spins and their classical model simultaneously have the dynamical symmetry generated by the subalgebra used to define the SDF, and if the interaction is such then the quantum system does not generate quantumness. Otherwise if the dynamical symmetry
     is broken the quantum system generates quantumness. A particular feature of this example is that if the dynamical symmetry that corresponding to the SDF is broken then the classical model is not integrable. Of course, there are classical Hamiltonian systems with potential interaction that breaks the symmetry used to define the SDF, but that are nevertheless completely integrable. An  example is provided next.

\subsubsection{Coupled oscillators: $h_4\oplus h_4$ dynamical algebra}

The dynamical algebra and the subalgebra chain used to specify the SDF are the same as in the example 1 of section 2.2.
Therefore the system has two SDF corresponding to the algebra elements given by the Casimir elements $n_1$ and $n_2$ of the subalgebra.
Coherent states $|p,q\rangle$ are again separable and given by the product of coherent states $|p,q\rangle\equiv |p_1,q_1\rangle|p_2,q_2\rangle$
 for each of the SDF.

Consider the quantum Hamiltonian
\begin{equation}
\hat H=\sum_{i=1,2} \frac{1}{2}[\hat p_i^2+\hat q_i^2] +\mu_1 \hat q_1^2\hat q_2-\frac{\mu_2}{3} \hat q_2^3
\end{equation}

Observe that $\hat q_i,\hat p_i$ denote the coordinate and the momentum operators, while $(q_i,p_i)$ are not their eigenvalues (which shall not occur) but are the parameters of the $h_4\oplus h_4$ coherent states.

Dynamics of the quantumness $\Delta_{h_4\oplus h_4}(\psi)=\sum_{i=1,2}\Delta_{\psi}\hat q_i\Delta_{\psi}\hat p_i$ is computed  staring from
 a coherent state $|p,q\rangle$.
Figure 3 demonstrate that when $(\mu_1=1,\mu_2=1)$ then $\Delta_{h_4\oplus h_4}(\psi(t))\neq const$ and when $(\mu_1= 0,\mu_2=0)$ or $(\mu_1= 0,\mu_2=1)$
  then $\Delta_{h_4\oplus h_4}(\psi(t))= const=0.5$ is minimal all the time. Thus, the example suggests that  as long as there is no interaction between the two SDF there isc
   no generation of quantumness, even if the dynamics of the separated SDF is governed by a non-quadratic Hamiltonian. On the other hand, the quantumness is
    generated if there is specific interaction between the SDF. Observe that the interaction in (14) is such that the system with interaction is not symmetric under the subgroup generated by the subalgebra used to define SDF. In other words, the Casimir operators $\hat n_1$ and $\hat n_2$,
      that are used to identify the two SDF do not commute with the interaction.

       \begin{figure*}
\centering
\includegraphics[width=1\textwidth]{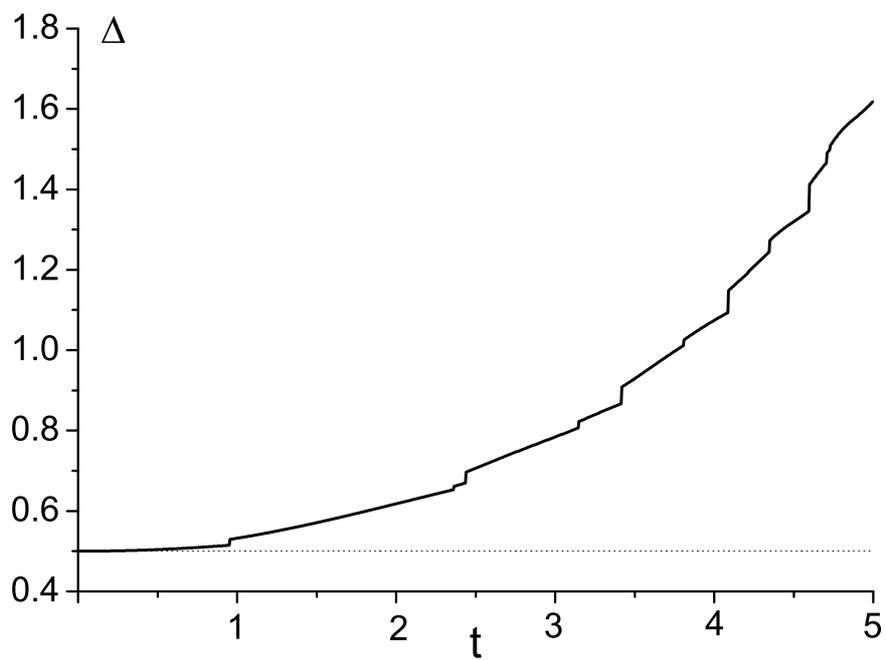}
\caption{ Illustrates dynamics of the quantumness $\Delta_{h_4\oplus h_4}$ with (14) when $\mu_1=0,\mu_2=1$ (dotted) and $(\mu_1=\mu_2=1)$ (full line).}
\end{figure*}

  {\it Dynamics of the classical model}

  The Classical model is given by
  \begin{equation}
  H_{p,q}=\sum_{i=1,2}\frac{1}{2}[ p_i^2+ q_i^2]+\langle p,q| \hat \mu_1 q_1^2\hat q_2+\frac{\mu_2}{3}\hat q_2^3|p,q\rangle
\end{equation}

Using the general formula for a Hamiltonian of the form $\hat H=\sum_i\hat p_i^2/2m_i+\hat V(q_1,q_2)$
\begin{equation}\label{e:H_sum}
H_{p,q}={\sum_i p_i^2/ 2m_i}+V(x)+\sum_{k=1}^{\infty} {1\over 2^k k!}{\hbar^k V^{(2k)}(x)\over (2m\omega)^k},
\end{equation}
where $V^{2k}(x)$ denotes the sum of derivatives of order $2k$ of the potential $V(x)\equiv V(q_1,q_2)$, we obtain the explicit form
 of the Hamilton's function for the classical model
\begin{eqnarray}
H(q_1,q_2,p_1,p_2)&=&\frac{1}{2}[p_1^2+p_2^2+q_1^2+q_2^2]+\mu_1 q_1^2 q_2-\frac{\mu_2}{3} q_2^3\nonumber\\
&+&\frac{\hbar}{2}[1+\mu_1(q_1+q_2)-\mu_2 q_2].
\end{eqnarray}

 Many typical orbits are computed  and it is demonstrated that when $(\mu_1=0,\mu_2=0)$ or $(\mu_2=1,\mu_1=0)$ then  all orbits are periodic or quasi-periodic and when $(\mu_2=1,\mu_1=1)$ then there are irregular orbits. The results are illustrated in figure 4. by plotting $q_1(t)$ component of a periodic (fig.4a) and chaotic (fig. 4b) orbits.
  Other interesting cases of the parameter $(\mu_1,\mu_2)$ values will be discussed shortly.
   \begin{figure*}
\centering
\includegraphics[width=1\textwidth]{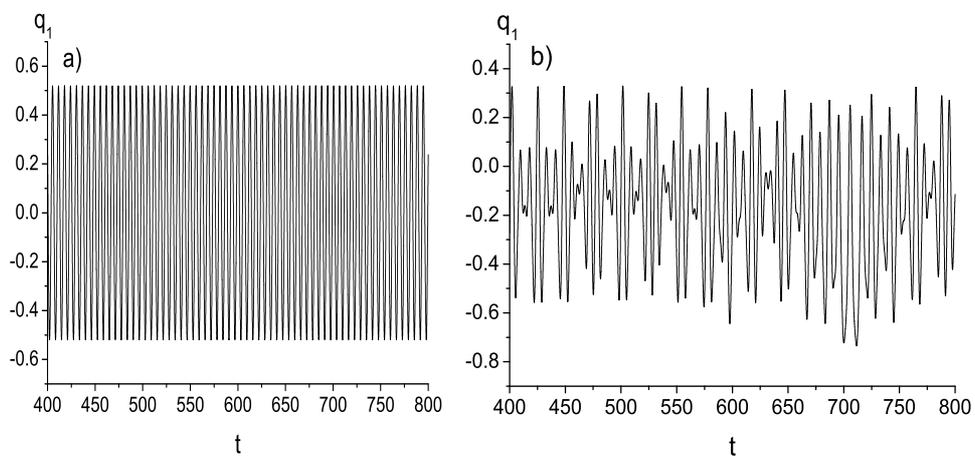}
\caption{ Illustrates dynamics of the classical model with the Hamiltonian (17) for a) $(\mu_1=0,\mu_2=1)$ and b) $(\mu_1=1,\mu_2=1)$. }
\end{figure*}

  {\it Dynamics of the classical model in the macro-limit}

 The Hamilton's function of a classical system is obtained from the classical model in the macro-limit $\hbar\to 0$
   \begin{equation}
   H_{cl}=\frac{1}{2}[p_1^2+p_2^2+q_1^2+q_2^2]+ \mu_1 q_1^2 q_2-\frac{\mu_2}{3} q_2^3.
\end{equation}
Orbits are computed  and it is demonstrated that for $(\mu_1=0)$ (and arbitrary $\mu_2$) all orbits are periodic or quasi-periodic and when $(\mu_1=\mu_2 =1)$
  then some orbits are irregular. In fact, it is known that the classical system, known as the Henon-Hiles model \cite{HH}, is completely integrable for
   the following cases of the parameter values: a) $(\mu_1/\mu_2=0)$; b) $\mu_1/\mu_2=-1$ and c) $\mu_1/\mu_2=-1/6$.
   Observe that the classical model (17) for the parameter values a) is also integrable (separable) with some bounded orbits, and for these parameter
    values the quantum system does not generate quatumness.
    The last two integrable cases are of special interest for the comparison with the quantum model.
 The classical model with $\mu_1/\mu_2=-1$ is integrable, but the SDF are not separated and $H_4\otimes H_4$ is not a symmetry of the classical model. For these parameter values the quantum model generates quantumness. So, it is not just the complete integrability of the classical system which is
  enough to imply the lack of quantumness generation by the quantized system.

 We have also tested a system of two harmonic oscillators with the simplest interaction $\hat V=\hat q_1 \hat q_2$, in which case the classical system and the classical model are the same. The classical system is integrable but the corresponding quantum system generates quantumness (please see fig. 5).

   \begin{figure*}
\centering
\includegraphics[width=1\textwidth]{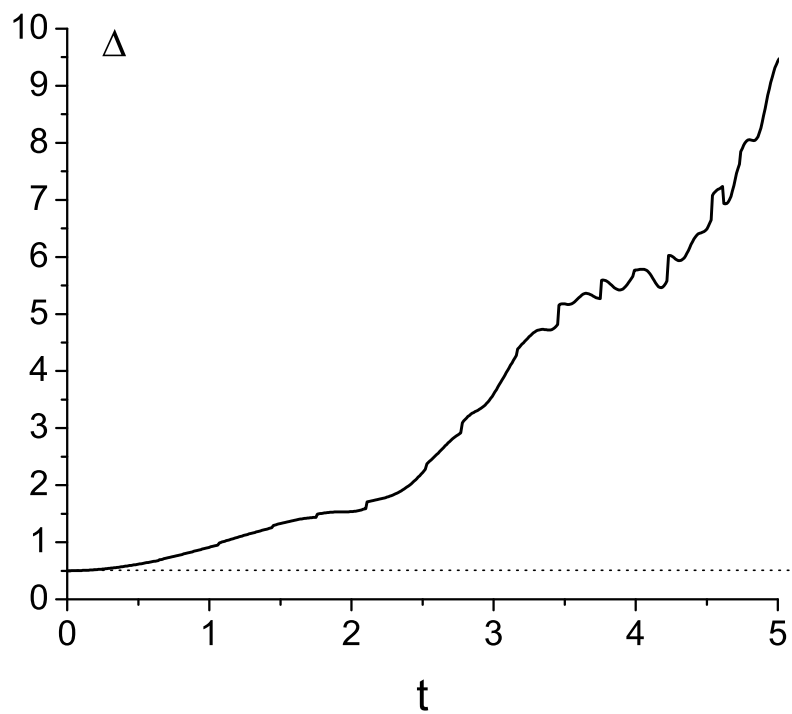}
\caption{ Illustrates dynamics of the quantumness $\Delta_{h_4\oplus h_4}$ with the interaction between the harmonic oscillators of the form $\hat q_1 \hat q_2$. Dotted line illustrates the constant $\Delta_{h_4\oplus h_4}$ with the Hamiltonian (19).}
\end{figure*}

  The conclusion of all computations with the example (14) and its classical model (17) is as follows. If the classical model is not integrable the quantum system generates quantumness.
  However, the quantum system might generate quantumness even if the corresponding classical model is completely integrable. In the considered examples, this happens if the interaction between SDF  is such that the subgroup generated by the subalgebra chain used to define the SDF is not a dynamical symmetry. The classical model is completely integrable, so there exist the corresponding action-angle variables. However, the action variables and the variables corresponding to the SDF are related by a nonlinear transformation. Therefore, the commutation relations between SDF variables and the quantum Hamiltonian do not reproduce the corresponding Poisson brackets and the quantum SDF variables do not generate  a dynamical symmetry.

   The last observation lead us to considered a system of coupled harmonic oscillators such that the subgroup generated by the subalgebra chain used to define the  SDF is a dynamical symmetry. The Hamiltonian of the quantum system is
  \begin{equation}
\hat H=\sum_{i=1,2} \frac{1}{2}[\hat p_i^2+\hat q_i^2] +\frac{1}{4}[\hat p_1^2+\hat q_1^2][[\hat p_2^2+\hat q_2^2],
\end{equation}

 In this case the quantum system does not generate quantumness, despite the interaction between SDF. The classical model and the classical system differ by a constant only, and
 are completely integrable.  More importantly they are in a subclass of integrable systems such that the Hamiltonian is expressed
 in terms of only the Casimir elements used to define SDF  (the action variables of the harmonic oscillators). The reason for the lack of quantumness generation is by now quite clear: the subgroup generated by the subalgebra chain used to define the  SDF is a dynamical symmetry.

Other examples that we have treated include systems with dynamical algebras $g=su(2)\oplus h_4$ and $g=su(3)$. These also support the general conjecture formulated in the next subsection. We shall briefly comment the case of systems providing realizations of  $su(3)$ dynamical algebra, because they provide the opportunity to illustrate few interesting features.

\subsubsection{ A simple system with more than one SDF: $su(3)$ dynamical
algebra}

The example of $su(3)$ dynamical algebra is used to illustrate the
systems with more than one SDF which nevertheless can not be
considered as composed of component systems with fewer number of
SDF because the irrep space of states does not have the
corresponding tensor product structure. Thus, the system should not be considered as simple in the sense that it is not
 composed of simpler systems.  The example will also
illustrate another important fact, namely the fact that the number
of SDF might depend on the particular irrep that is carried by
 the system's Hilbert space. However, the general conjecture about NDDF,
  concerning the relation between the dynamical symmetry, generation of quantumness and dynamics of the classical
   models seems to be confirmed also by the $su(3)$ examples.

The $su(3)$ Lie algebra has rank $2$ and dimension $8$. The basic commutation relations between the
 generators $E_{i,j}, \>i,j=1,2,3$ which are
 not independent are: $[E_{ij},E_{kl}]=\delta_{jk}E_{il}-\delta_{il}E_{kj}$ and can be realized
 in terms of bosonic creation and annihilation operators of three modes as follows: $E_{i,j}=a_i^{\dag}a_j,\>i,j=1,2,3$.
 The eight independent  hermitian generators are given by:
 $X_1=(a_1^{\dag}a_1 -a_2^{\dag}a_2);\>X_2=(a_1^{\dag}a_1
 -a_2^{\dag}a_2-2a_3^{\dag}a_3);\>Y_k=i(a_k^{\dag}a_j-a_j^{\dag}a_k);\>Z_k=(a_k^{\dag}a_j-a_j^{\dag}a_k),\>
 k=1,2,3,\> j=k+1 \>({\rm mod} 3)$. These will be  used in the formula
 (5) for the level of $su(3)$-quantumnes in a particular system with the corresponding SDF.

In order to determine the number of SDF we need to find the number
of nonfully degenerate operators in any  particular chain of
subalgebras. We shall use the subalgebra chain: $su(3)\supset
su(2)\oplus u(1)\supset u(1)$ with five Casimir operators usually
denoted by $C_2,C_3,Y,T^2,T_z$. $C_2$ and $C_3$ are the Casimir
operators of the su(3) itself, $T^2$ and $T_z$ are the Casimir
operators of $su(2)$ and u(1) and $Y$ corresponds to $u(1)$.
 Thus, in general there are three nonfully
degenerate operator and consequently a system with $su(3)$ algebra
has three
 SDF. However, the system is also characterized by its Hilbert space i.e. by a particular irrep and for some irrep
 all three DF might not
  be independent.

All irreps of the $su(3)$ algebra can be labeled by their highest
weight: $\Lambda=\lambda_1f_1+\lambda_2 f_2$ where $f_1$ and $f_2$
are the highest
 weights of the two fundamental representations: $(1,0)$ and $(0,1)$.  The fully symmetric representations correspond to
 $\lambda_1=0$ or $\lambda_2=0$.
In the fully symmetric representation the operators $T^2$ and $Y$
are not independent and thus in this case the number of SDF is
just 2. A system with such SDF has the $su(3)$ dynamical symmetry if its Hamiltonian is expressed in terms of $T^2$ and $T_z$  or $Y,T^2$ and $T_z$ in the two or
  three degrees of freedom cases.

The coherent states of the $SU(3)$ dynamical group are obtained as in the general case, using the highest weight vector as the reference state $|\psi_0\rangle$.
 In the general case the coherent states are parameterized by the six dimensional manifold: $SU(3)/U(1)\otimes U(1)$ and in the case of
 the fully symmetric irrep with two SDF by the four dimensional $SU(3)/U(2)$. As usual the coherent states are of the form
$|\Lambda,\alpha>= D(\alpha)|\psi_0\rangle$, where $\Lambda$ is fixed by the irrep and $\alpha$ indexes different coherent states. According to the adopted definition the coherent states have minimal quantumness. The minimal quantumness is preserved by systems with Hamiltonians linear in terms of the algebra generators used to define the corresponding SDF.
 Dynamics od $su(3)$ quantumness with Hamiltonians nonlinear in the generators corresponding to SDF  is illustrated in the following example.

Consider the system of $N$ particles with three possible
$N_d$-degenerate energy levels. The following Hamiltonian for such
a system is known
 as the Lipkin model:
\begin{equation}
H=\sum_{i=1}^3\omega_i E_{ii}-\mu\sum_{i\neq j}^3E_{i,j}^2
\end{equation}
where $E_{ij}$ satisfy $su(3)$ commutation relations.   Dynamical symmetries of systems with such
Hamiltonians and dynamics of the corresponding classical models were studied in
\cite{Cheng2}.
 When $N\leq N_d$ the Hilbert space of the system is the carrier space of the fully-symmetric
 irrep and the system has two SDF. If $\mu=0$ there is the dynamical symmetry corresponding to
  $su(3)\supset su(2)\oplus u(1)\supset u(1)\times u(1)$,
 and the system does not generate quantumness .
 For $\mu\neq 0\neq \omega_i$ the dynamical symmetry corresponding to the SDF is broken and the system generates
    quantumness  with respect to the relevant SDF. This is illustrated in figure 6.

       \begin{figure*}
\centering
\includegraphics[width=1\textwidth]{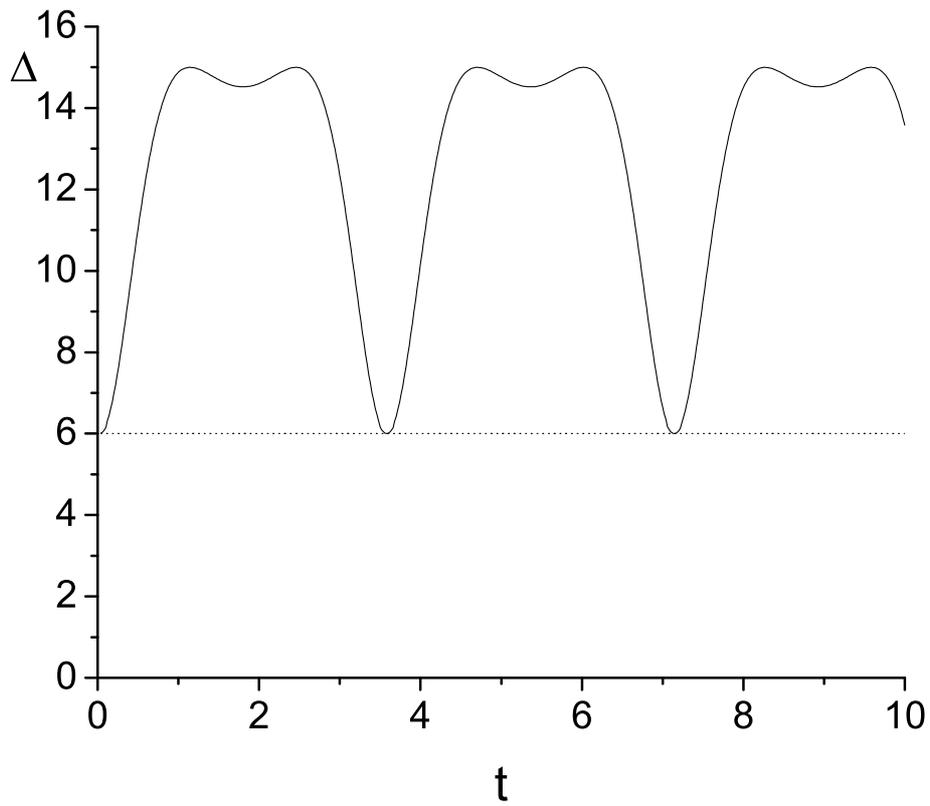}
\caption{Illustrates dynamics of the quantumness $\Delta_{su(3)}$ with
the Hamiltonian (20), starting from an SU(3) coherent states in the completely
symmetric irrep. Full line corresponds to $\mu=1/6,\> \omega_i=1$
 and dotted to the symmetric case $\mu=0,\> \omega_i=1$.}
\end{figure*}

{\bf Remark} Properties of quantum systems that indicate if the system is obtained by quantization of an  integrable or non-integrable classical system have been studied intensively in the past ( Please see for example \cite{q_chaos}). For example, such properties are distributions of spectral levels and dynamics of quantum phase-space distributions including their zeros. However, contrary to the studies of classical dynamics where qualitative properties of orbits are central, in these studies of quantum systems properties of state-vector orbits are usually not considered. The reason for this is that orbits of any quantum system are either periodic or quasi-periodic, and no direct qualitative comparison with typically chaotic classical orbits is possible. On the other hand, the notion of DDF is directly related to the dynamics of generic orbits and is relevant for classical as well as quantum systems. Also, generation of quantumness is a property of quantum dynamics that directly reflects relevant properties of the corresponding classical model.
Furthermore, classical models exist for a much larger class of quantum systems than those obtained by quantization of classical systems.

\subsection{Formulation of the main conjecture}

Based on the above examples we formulate the following conjecture for quantum systems with a dynamical Lie algebra and the corresponding classical models

{\it Conjecture}

 If the dynamics preserves dynamical variables that correspond to the
 non-degenerate Casimir elements of the subalgebra chain used to define
 the SDF than NSDF=NDDF for the quantum as well as the classical systems. In this case the quantum system does not generate quantumness and the quantum system and the classical model are given in terms of the
  Casimir elements corresponding to the SDF.

We have seen examples of classical models which are integrable but not in the form of the previous conjecture. In this case, the corresponding
 quantum system generates quantumness with respect to the considered SDF, and the NDDF $>$ NSDF for the quantum system. On the other hand, in general for the classical integrable systems NDDF= NSDF.

In the case of generation of quantumness and non-integrability of the classical model we know that
 NDDF$>$NSDF for the quantum and the classical case, but we can not make any prediction in general concerning the relation between NDDF of a quantum system and the corresponding classical model.

 \section{Summary}

We have discussed three different types of degrees of freedom (DF): formal, structural and dynamical, that are meaningful and useful in descriptions of quantum and classical systems. The formal (FDF) and the dynamical (DDF) are related to the dynamics; the formal to the type of evolution equation of the class of systems and the dynamical to the relevant properties of a particular system dynamics. On the other hand, the structural DF (SDF) represent what is commonly understood by DF, and describe structural properties  of a system, and not its dynamics. SDF have been defined quite generally for systems
 with basic variables forming a realization of a Lie algebra. An appropriate Lie-algebra uniquely determines the number of SDF and a particular chain of subalgebras determines which are the SDF. Considerations of properties of typical orbits of generic Hamiltonian systems suggest that a notion of
   dynamical DF (DDF), different in number and type from SDF, is useful and important. Similarly, considerations of entanglement dynamics of typical quantum systems also suggest an analogous notion of DDF. We defined the notion of DDF, for classical as well as quantum systems, as the dimension of
    dynamically irreducible and invariant manifold generic for a given system. The number of DDF is generically larger than the number of SDF.

    We then defined the notion of quantumness of a state with respect to some SDF. It turns out that the Lie algebraic generalized coherent states are the states with minimal quantumness. The generalized coherent states are also used to define an appropriate classical model of a quantum system,
     with the same number and type of SDF and coarse-grained dynamics.

  Our next task was to examine the relation between the numbers of DDF and SDF for a quantum system and its classical model. To this end we have studied relations between generation of quantumness by a quantum system and the dynamics of its classical model. Analyzes of relevant examples
   suggested the general conjecture that: A quantum system with SDF determined by an algebra and its particular chain of subalgebras does not generate quantumness with respect to the SDF if and only if the subalgebras used to define the SDF generate dynamical symmetries. If the quantumness can not be dynamically generated than
 the Hamiltonians of the quantum system and its classical model is necessarily  expressed solely in terms of the
  Casimir elements used to define the SDF. The classical systems with this property are certainly completely integrable. Of course, there are
  classical completely integrable systems with Hamiltonian depending on variables other than the Casimir elements related to the SDF.
  We have demonstrated examples of quantum systems with such integrable classical models which do generate quantumness with respect to the SDF. This shows that the above conjecture cannot be extended to include all completely integrable classical models.

  \vskip 0.5cm

 {\bf Acknowledgements} This work was supported in part by the Ministry of Education and
Science of the Republic of Serbia, under project No. $171017$

\end{document}